\documentclass[twocolumn,showpacs,preprintnumbers,amsmath,amssymb]{revtex4}
\usepackage{graphicx}% Include figure files
\usepackage{dcolumn}% Align table columns on decimal point
\usepackage{bm}% bold math

\begin{document}

\title{Metallic and Insulating behaviour in p-SiGe\\
at $\nu$ = 3/2}

\author{P.T. Coleridge, R.L. Williams, J. Lapointe and P.
Zawadzki}

\affiliation{Institute for Microstructural Sciences\\
 National Research Council\\ Ottawa, Ontario,K1A 0R6, Canada}

\begin{abstract}
Shubnikov-de Haas data is presented for a p-SiGe sample
exhibiting strongly insulating behaviour at $\nu$  = 3/2. In
addition to the fixed points defining a high field metal-
insulator transition into this phase separate fixed points can
also be identified for the $\nu = 3 \rightarrow 2 $ and $2
\rightarrow 1$ Integer quantum  Hall transitions. Another
feature of the data, that the Hall resistivity approaches zero
in the insulating phase, indicates it is not a re-entrant Hall
insulator. The behaviour is explained in terms of the strong
exchange interactions. At integer filling factors these cause
the $0 \uparrow$ and $1 \downarrow$ Landau levels to cross and
be well separated but at non-integer values of $\nu$ screening
reduces exchange effects and causes the levels to stick
together. It is suggested the insulating behaviour, and high
field metal/insulator transition, is a consequence of the
strong exchange interactions.
\end{abstract}

\maketitle

\section{Introduction}

     The appearance, in p-SiGe, of insulating behaviour between
filling factors $\nu$ = 1 and 2 has been known for several
years \cite{fang92,dorozhkin95,dunford96,coleridge97a,lin98}. 
Although not always present it has been observed by several
groups in samples from more than one source. Similar behaviour
has also been seen recently \cite{sakr01} near filling factors
5/2 and 4. 

     There is, as yet, no generally accepted explanation for
this phenomenon. Two possibilities are that the insulating
phase (IP) might be attributed to a premature and re-entrant
Hall insulating state \cite{lin98} or that it results from a
combination of the unusual ordering of the Landau levels and a
long-range modulation of the potential \cite{dorozhkin95}. An
alternative suggestion \cite{coleridge97a} is that the IP
appears when the large, exchange enhanced, Zeeman splitting 
causes adjacent Landau levels to cross and could be associated
with an unusual spin texture or domain structure.  Results are
presented here to support this view. It is shown (a) that the
behaviour of the Hall resistivity in the IP is not consistent
with a Hall insulating state and (b) that although no exotic
spin texture states are expected \cite{giuliani85} at $\nu=2$
this is not necessarily the case for non-integer filling
factors, particularly when disorder and screening are taken
into account.

\section{Experiment}

     Figure 1 shows details of the temperature dependence of
the Shubnikov-de Haas (SdH) oscillations in a p-SiGe sample
\cite{coleridge97b} with a density of $1.7 \times
10^{15}$m$^{-2}$ and mobility of 1.65 m$^2$(Vs)$^{-1}$.  The IP
which appears at $\nu$ = 3/2 is characterised by fixed points
with $\rho_{xx} \sim h/e^2$ (see also figure 2a). They indicate
a metal-insulator transition with scaling behaviour \cite{ref7}
similar to that observed for transitions into the high field
($\nu > 1$) Hall insulating state. 
      Separate, well defined, fixed points associated with
Integer Quantum Hall Effect (IQHE) transitions can also be
observed.  These are shifted somewhat from the expected
magnetic fields but the critical values (in units of $e^2/h$)
of $\sigma_{xx}^c$ = 0.65 and  $\sigma_{xy}^c$ = 2.7 for the 3
$\rightarrow$ 2  transition and $\sigma_{xx}^c$ = 0.59,
$\sigma_{xy}^c$ = 1.6 for the 2 $\rightarrow 1$  transition,
are close to the expected values of (0.5,2.5)  and (0.5,1.5)
respectively. Near  B = 3 tesla the increase of $\rho_{xx}$
with temperature is metallic-like, mirroring the behaviour
around 4 tesla.  This is interpreted as the precursor to
another IP which is suppressed by the onset of the $\nu$ = 2
IQHE plateau before becoming fully developed. (In a lower
density sample this peak becomes fully insulating
\cite{sakr01}).

A feature of the data is the behaviour of $\rho_{xy}$ through
the IP. This is not an easy measurement, because any sample
inhomogeneity produces spurious $\rho_{xx}$ terms which can be
an order of magnitude larger than the genuine signal, but when
these are removed by averaging data with normal and reversed
magnetic fields, it can be seen that the Hall resistance tends
to vanish as the insulating behaviour strengthens. This is a
general feature of the IP: figure 2b shows measurements in a
slightly lower density sample and similar trends can also be
identified in other data in the literature
\cite{dunford96,lin98}.

     A vanishing Hall resistance is incompatible with the
$\nu$= 3/2 IP being a re-entrant Hall insulating state. The
Hall insulator \cite{ref9} appears when the Fermi energy lies
in the last Landau level: in the IQHE regime it is
characterised by a Hall conductance that vanishes as
$\sigma_{xx}^2$ so the Hall resistivity remains approximately
quantised at $h/e^2$, in the fractional QHE regime it increases
approximately as $B/n_s e$ (where n$_s$ is the density) with
FQHE plateaus appearing over limited field ranges. The
vanishing of
$\rho_{xy}$ therefore indicates the insulating phase is neither
an IQHE nor a FQHE Hall insulator. 
   
     Coulomb interaction and exchange effects are large in
these p-SiGe samples. For example, for that shown in figure 1,
at $\nu$=2, the exchange energy [$(\pi/2)^{1/2} (e^2/\kappa
l_m)$ where $l_m$ is the magnetic length] is 10meV,
approximately five times larger than the cyclotron energy
($\hbar\omega_c$) of 1.9 meV. The (bare) spin splitting in p-
SiGe is relatively large, of order $\hbar\omega_c/2$, and is
suficiently enhanced by exchange that the $0\uparrow$ and
$1\downarrow$ Landau levels cross and the system becomes
ferromagnetically polarised at $\nu$=2 (see figure 3).
Activation measurements in other p-SiGe samples
\cite{coleridge97a} confirm this is so. At integer filling
factors, when the Fermi energy lies in localised states between
Landau levels, screening is unimportant, but for non-integer
filling factors, when the density of extended states at the
Fermi energy is large, it can significantly reduce the exchange
enhancement of the spin splitting. A model
calculation is presented that explores this situation.

\section{Model calculation}

     The calculation uses the formalism given by Yarlagadda
\cite{yarlagadda91} with the screening of the bare coulomb
potential ($V_q$) given by

\begin{equation}
 \epsilon(q) = 
     1 + V_q (D_0^{\uparrow} + D_1^{\downarrow} +
\Pi^{interLL}) 
\end{equation}
where $D_N^{\sigma}$ is the density of states at the Fermi
level of the $N\sigma$ Landau level and $\Pi^{interLL}$ is an
inter-Landau level term. For any significant density of states
at the Fermi energy this last term is relatively small and will
be ignored.  The separation of the $1\downarrow$ and
$0\uparrow$ Landau levels is then given by

\begin{equation}
     \Sigma_1^{\downarrow} - \Sigma_0^{\uparrow}
               = \hbar \omega_c - g \mu_B B + \sum_q E_{ex}(q)
\end{equation}
where
\begin{equation} 
          E_{ex}(q)  = \frac{V_q}{\epsilon(q)} 
  (J_{00}^2 n_0^{\uparrow} - J_{11}^2 n_1^{\downarrow} -
     J_{10}^2 n_0^{\downarrow} )    
\end{equation}
with  $J_{NM}(ql_m)$ the coupling between the N and M Landau
levels and  $n_0^{\uparrow}$ etc partial filling factors. One
can also calculate the total (Hartree Fock) energy of the
system (E$^{HF}$) as the integral over occupied states of
$E_{KE} -\frac{1}{2} E_{ex}$ where $E_{KE}$ is the kinetic
energy.

     Realistic parameters for the sample shown in figure 1 were
used: an effective mass of 0.2$m_e$, a bare Zeeman splitting of
0.65 $\hbar \omega_c$ and a (Gaussian) Landau level broadening
given, according to the self consistent Born approximation, by
$(e \hbar^2 \omega_c/2 \pi m^{\ast} \mu_q)^{1/2}$ where the
(measured) quantum mobility $\mu_q$ is 1.7m$^2$(Vs)$^{-1}$.

     Figure 4 shows the separation between the
$\Sigma_1^{\downarrow}$ and $\Sigma_0^{\uparrow}$ Landau levels
as a function of $n_0^{\uparrow}$, determined (a) from eqns. 2
and 3 and (b) from the constraint that
$n_0^{\downarrow}+ n_0^{\uparrow} +n_1^{\downarrow} = \nu$. It
is assumed the lowest, $0^{\downarrow}$, level always remains
full. Valid solutions occur at the intersection of these two
curves. At $\nu$=2, the calculation without screening (see
figure 4a) shows an unstable, degenerate solution and two
stable solutions corresponding to either  a ferromagnetic or
paramagnetic ordering of the Landau levels (the former with the
lowest energy). This essentially reproduces the result of
Giuliani and Quinn \cite{giuliani85} who also showed that for
non-degenerate Landau levels one of these stable solutions will
pre-empt any spin-density-like state.

     When screening is included the ferromagnetic configuration
becomes even more strongly favoured but the other solutions are
significantly altered, in particular the paramagnetic solution
now involves degenerate Landau levels and there is an increase
in the total energy.  When the filling factor moves away from
two more significant differences appear (see figure 4b). The
density of states at the Fermi energy is always non-zero,
screening becomes important and the exchange
enhancement of the spin splitting is reduced. Under these
conditions only one, degenerate, solution remains and the
ferromagnetically polarised state can no longer exist.

     Figure 5 shows the Landau level separation as a function
of filling factor. The ferromagnetically polarised state
appears only at integer filling factors, when the screening is
small. 
Elsewhere, the levels stay close together and are degenerate
with a predominantly paramagnetic alignment. The transition
into the ferromagnetically polarised state is sharp near
$\nu$=2 but much smoother near $\nu$= 1 and 3.

\section{Discussion}

     The model calculation can explain the existence of both a
high field metal/insulator transition and separate, well
defined, IQHE transitions.  Near integer filling factors, the
exchange interaction ensures the Landau levels are well
separated and the IQHE transitions can then occur in the tail
of an isolated Landau level. Away from integer filling factors,
screening means the Landau levels overlap which correlates with
the appearance of the insulating (or metallic) behaviour.
Separate metal/insulator and IQHE transitions occur because the
Landau levels realign as the filling factor changes. In some
circumstances, for example in lower mobility samples or at
higher temperatures, this realignment may not occur and the
high field IP and IQHE transitions will then merge and no
longer be distinguishable.

     Away from integer filling factors the only allowed
solution corresponds to degenerate Landau levels (at least for
a homogeneous system). Although the Hartree-Fock energy in
figure 4b appears relatively flat as a function of the relative
populations of the two levels, a detailed examination shows
there could be a small lowering of the energy if the system
were to break up into domains. While the exact nature of such
a state (or some other kind of spin texture) is uncertain such
a mechanism would appear to be a candidate for the cause of the
insulating behaviour.

     In high mobility, GaAs based, 2DEG samples, coulomb
interaction effects manifest themselves as fractional quantum
Hall effect features. Because of the difference in effective
masses (0.2$m_e$ compared with 0.067$m_e$) and the ratio of the
transport to quantum lifetimes (approximately one in p-SiGe
compared with 20 or so in n-GaAs 2DEGs) the Landau level
broadening in the p-SiGe sample investigated here is very
similiar to that of a GaAs based 2DEG with a mobility of over
200 m$^2$(Vs)$^{-1}$. The exchange energy depends only on
magnetic field so the absence of fractional quantum Hall
features is a little puzzling. The essential difference is that
in n-GaAs heterostructures higher Landau levels are well
removed in energy, typically by several times the interaction
energy, whereas in p-SiGe there are two degenerate Landau
levels at the Fermi energy. This is consistent with the
insulating phase (and associated high field metal-insulator
transition) being a direct consequence of strong exchange
interactions in degenerate Landau levels suppressing more usual
FQHE behaviour.

     The model calculation does not include a number of factors
such as: the finite thickness of the hole gas, correlation
effects, the existence of a mobility edge and the contribution
of inter-Landau level screening but the result, that screening
causes the Landau levels to stick together, is robust.  A
negative feed-back mechanism (relating the exchange splitting
of the Landau levels to the screening) ensures that the
separation of the levels is small and relatively independent of
exact values of the parameters, provided only that the
{\it unscreened} exchange energy is large compared with the
bare separation and that screening significantly reduces this
exchange energy. 

     In electron systems Landau level crossings occur and can
be tuned with parallel magnetic fields \cite{ref10}.  They are
identified by `spikes' in the Shubnikov-de Haas data with, in
some cases, hysteresis. This is interpreted in terms of domain
formation in these quantum Hall Ising ferromagnets.  The
question of spin textures in such systems has recently been
studied theoretically by Brey and Tejedor \cite{brey01}. They
find that a spin-orbit interaction, which directly couples the
$0\uparrow$ and $1\downarrow$ states,  provides a means of
producing a gap in the energy spectrum of the charged
excitations that cross domain walls. It is therefore
interesting to note that in p-SiGe, where strain separates the
two degenerate heavy-hole states at the Brillouin zone centre,
there is a large spin-orbit coupling with the holes having
almost pure $\mid M_J \mid$ = 3/2 character. However, it should
be also noted these systems are not necessarily directly
equivalent to that discussed here which often involve
relatively weakly coupled 2DEGs, either in a rather wide single
quantum well or in two wells separated by a barrier.

\section{Conclusion}

     Fixed points defining both IQHE transitions and a high
field insulating phase can be separately identified in
Shubnikov-de Haas measurements in a p-SiGe sample. Further, it
is found that the Hall resistivity in the IP tends towards zero
as the temperature is lowered, which implies the IP is not a
re-entrant Hall insulator.

     A model calculation is presented which explains these two
types of behaviour in terms of the large exchange interaction
which is screened when the Fermi energy lies within a Landau
level but not when it lies in localised states between Landau
levels. Insulating behaviour (or more generally a metal-
insulator transition) co-incides with the overlap of degenerate
Landau levels overlap at the Fermi energy. It is suggested
therefore that the insulating behaviour could well be a direct
consequence some form of domain structure or other kind of spin
texture induced by a strong exchange interaction in degenerate
Landau levels.


\newpage

\begin{figure*}
\vspace*{10.0cm}  
%\begin{center}
% Put the name of your file in the next line and remove the %
%\includegraphics{a:smagfig1.eps}
\includegraphics{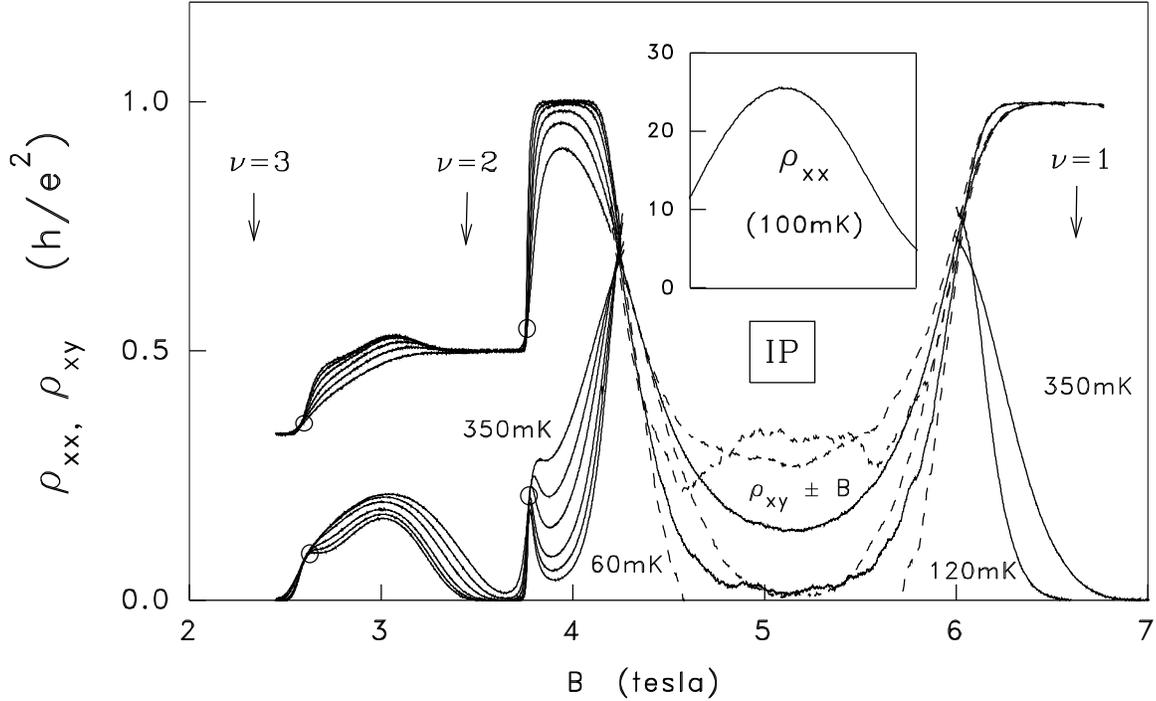}
%\end{center}
\caption{A composite figure showing Shubnikov-de Haas data for
the sample described in the text. The IP appears around 5
tesla. Below 4.3 tesla data is shown for six
temperatures between 60 and 350mK: note the fixed points
indicating IQHE transitions. Above 4.3 tesla $\rho_{xy}$ data
is shown at temperatures of 120 and 350mK for normal and
reversed magnetic fields (dashed) and for the average of these
(solid). The inset shows $\rho_{xx}$, at 100mK, determined from
a 2-terminal measurement.}
\label{figure1}
\end{figure*}

\begin{figure*}
\vspace*{8.0cm}
%\begin{center}
% Put the name of your file in the next line and remove the %
%\includegraphics{a:fig2.eps}
\includegraphics{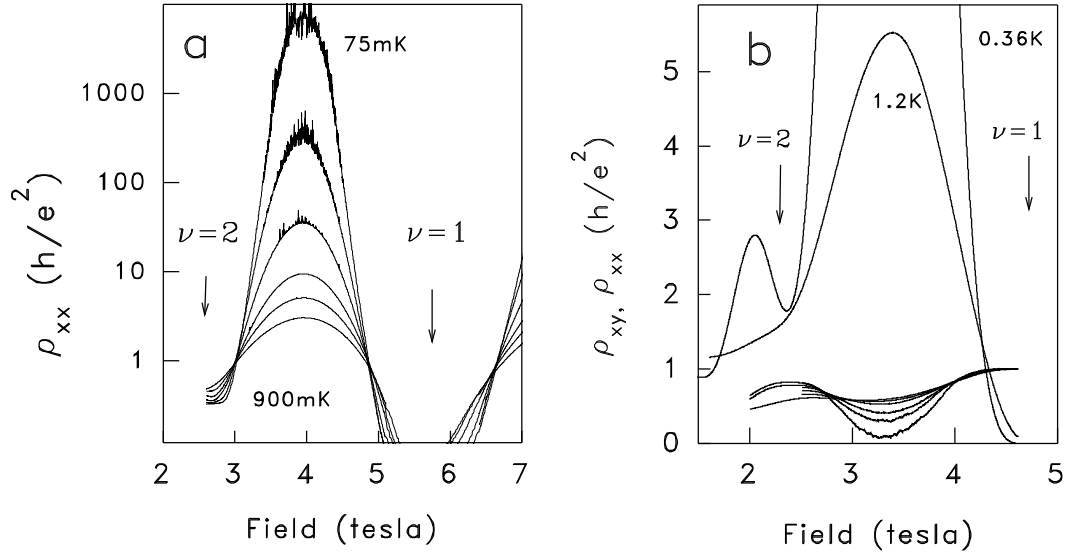}
%\end{center}
\caption{Shubnikov-de Haas oscillations from two p-SiGe samples
with densities of approximately 1.2$\times$10$^{15}$m$^{-2}$.
(a) $\rho_{xx}$, determined using a 2-terminal
measurement, for temperatures of 75, 120, 220, 400, 600 and
900mK. (b) Values of $\rho_{xx}$ at 0.36 and 1.20K and
$\rho_{xy}$ (averages for normal and reversed magnetic fields)
at 0.27, 0.36 ,0.45, 0.65, 0.90 and 1.20K. } 
\label{figure2}
\end{figure*}

\begin{figure*}
\vspace{7.0cm}  
%\begin{center}
% Put the name of your file in the next line and remove the %
%\includegraphics{a:smagfig1.eps}
\includegraphics{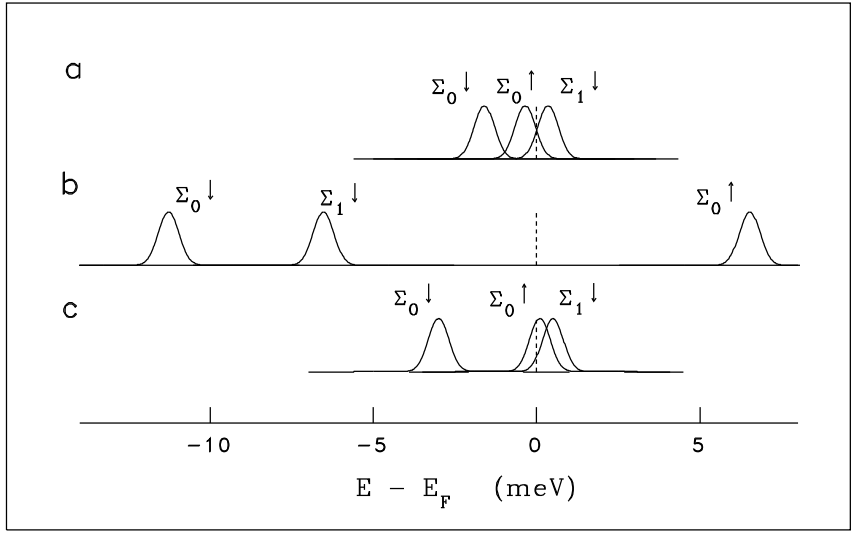} 
%\end{center}
\caption{Alignment of the three lowest Landau levels for sample
parameters appropriate to the data shown in figure 1. (a) At
$\nu$=2, using the bare value, $g \mu_B B$, for the spin
splitting. (b) At $\nu$ = 2 with exchange enhanced spin
splitting according to eqn. 2. (c) At $\nu$ = 3/2 with a
screened exchange term. } \label{figure3}
\end{figure*}

\begin{figure*}
\vspace{10.0cm}
%\begin{center}
%\includegraphics{a:fig4.eps}  
\includegraphics{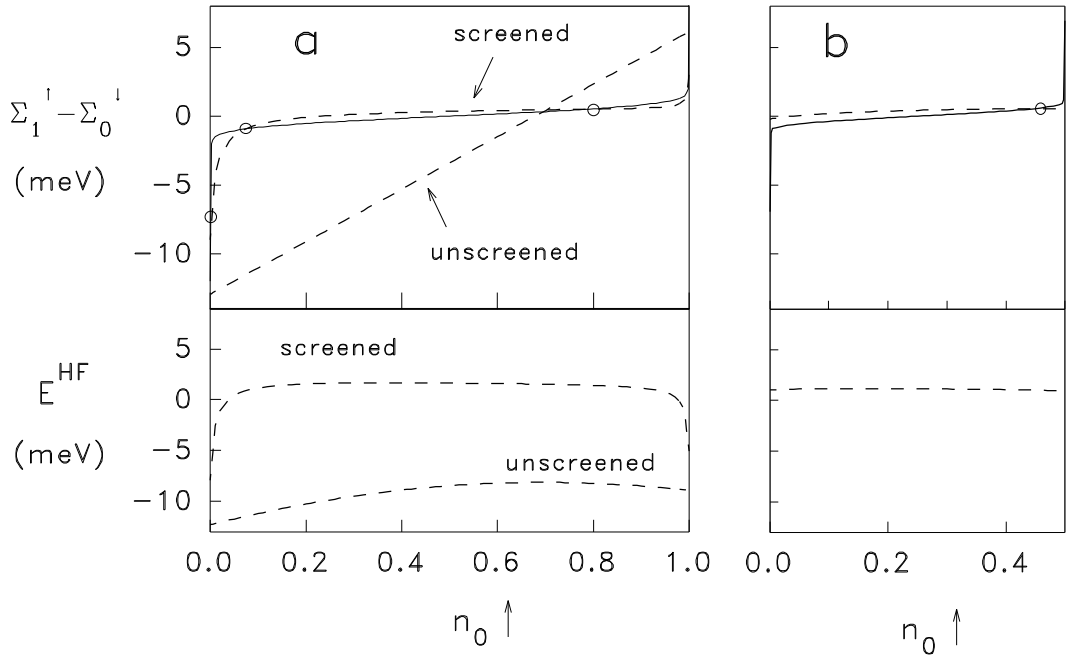} 
%\end{center}
\caption{Landau level spacing and total (Hartree-Fock) energy
as a function of partial filling factor $n_0^{\uparrow}$.
Dashed lines correspond to eqns. 2 and 3, solid lines to the
requirement that the sum of the partial filling factors must
equal $\nu$.  (a) at filling factor $\nu$ = 2,  with and
without screening. Circles show the three possible solutions
when screening is included. (b) at filling factor 1.5, in this
case only one solution is allowed.}
\label{figure4}
\end{figure*}

\begin{figure*} 
\vspace*{7.5cm}
%\begin{center}
\includegraphics{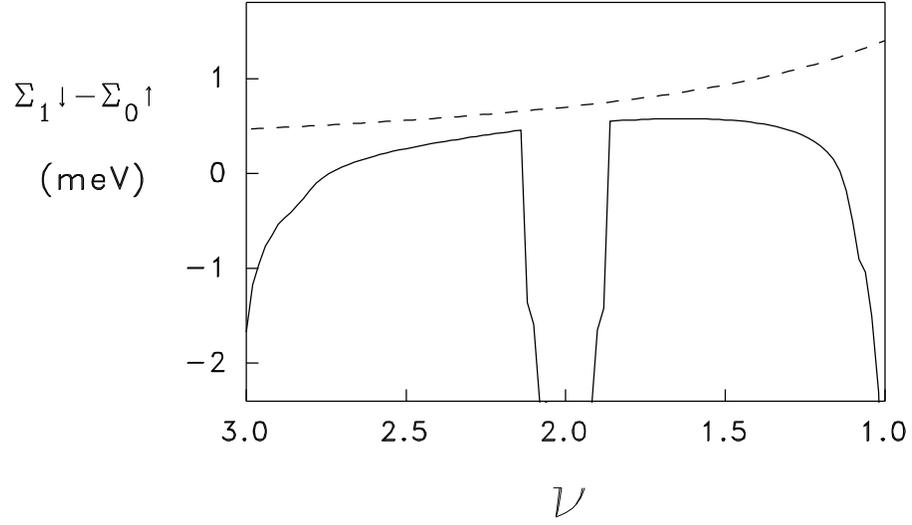} 
%\end{center}
\caption{Calculated Landau level separation as a function of
filling factor. Dashed curve shows the bare value ie $\hbar
\omega_c - g\mu_B B$. For comparison the width of a Landau
level (full-width at half maximum) is 0.74meV at $\nu$=2 and
varies as B$^{1/2}$.} 
\label{figure5}
\end{figure*}

\end{document}